\documentclass[prd,eqsecnum,floats,nopacs]{revtex4}
\usepackage{amsmath}
\usepackage{amssymb}
\usepackage{amsfonts}
\usepackage {graphicx}
\usepackage {graphics}
\usepackage {times}
\usepackage {txfonts}

\begin{document}

\title {Radiative effects in the  standard model extension}
\author{V. Ch. Zhukovsky}\email{zhukovsk@phys.msu.ru}
\author{ A. E. Lobanov}\email{lobanov@phys.msu.ru}
\author{ E. M. Murchikova}\email{murchikova@yahoo.com}
\affiliation {Department of Theoretical Physics, Moscow State
University, 119992 Moscow, Russia}

\begin{abstract}
The possibility of radiative effects induced by the  Lorentz and CPT
 non-invariant interaction term for fermions in the standard model
 extension is investigated. In particular, electron-positron photo-production
 and photon emission by electrons and positrons are studied. The
 rates of these processes are calculated in the Furry picture. It
 is demonstrated that the rates obtained in the framework of the model
 adopted strongly depend on the polarization states of the particles
 involved.
As a result, ultra-relativistic particles produced should occupy
states with a preferred spin orientation, i.e., photons have the
sign of polarization opposite to the sign of the effective
potential, while charged particle are preferably in the state with
the helicity coinciding with the sign of the effective potential.
This leads to evident spatial asymmetries which may have certain
consequences observable at high energy accelerators, and in
astrophysical and cosmological studies.
\end{abstract}


\maketitle

\section{Introduction}
Physical laws have been confirmed to be Lorentz invariant with
high accuracy in numerous experiments \cite{1}. Nevertheless, for
the last few years there have been made a number of assumptions
that these symmetries are only approximate.

The modern quantum field theoretical viewpoint admits the
possibility of Lorentz invariance breaking (and, as a consequence,
possible CPT invariance breaking in
the local field theory)  through a spontaneous
symmetry breaking mechanism. In other words, even though the
 underlying laws of nature have Lorentz and CPT
symmetries, the vacuum solution of the theory could spontaneously
 violate these symmetries.

The  Standard Model does not have dynamics necessary to cause
spontaneous Lorentz and CPT violation. However, the violation
mentioned above could occur in  more fundamental theories, such as
string theories, noncommutative geometry, etc., and the resulting
theory can be effectively described in the framework of the
standard model extension (SME) \cite{2}.

At present, there are numerous approaches to study various effects
of possible Lorentz violation. However, the SME is the most
consistent of all in describing theoretical implications of the
Lorentz violation hypothesis (for a recent review of SME, see,
e.g., \cite{new}).
It is in the framework of this
theory that calculations have been performed \cite{teor}, whose
results became the basis for experimental verification of Lorentz
invariance in the low energy range \cite{exp}.

It is surprising though that the high energy consequences
 of this theory  have not been studied as yet, whereas the
possible effects   in this range are expected to be more
substantial, than those at low energy.  Some estimates of the rates
of certain electromagnetic processes  have been
obtained only at the phenomenological level (see, e.g., \cite{1c} and
references therein). In the present publication, we attempted to partly
compensate for this lack of theoretical estimates.

The SME is fairly complicated even in the electromagnetic sector,
and hence we restricted ourselves to studying only those
consequences of the theory that are due to axial-vector
interaction of fermions with the constant background field
$b^{\mu}$  with the CPT-odd interaction term in the fermion
Lagrangian $\overline{\psi}b_{\mu}\gamma^{\mu}\gamma_5\psi$
(notations generally excepted in the framework of the SME
\cite{2}). This  kind of modification of QED in the fermion sector
does not influence the gauge invariance of the action and of the
equations of motion, but it does modify the dispersion relations
for Dirac spinors \cite{2}. The question about the possible
dynamical origin of this constant vector $b^{\mu}$ remains an
interesting task to be solved. One of the possibilities is that
the pseudovector field $b^{\mu}$  might be related to some
constant background torsion in the large scale Universe,
$b^{\mu}\sim\varepsilon^{\mu\nu\lambda\delta}T_{\nu\lambda\delta}$
\cite{6}. Moreover, such a CPT-odd term could be generated by
chiral fermions \cite{7}.

This model has been extensively employed in numerous publications
\cite{9,10,11,12,13,14}, where the Chern--Simons term generated by
the fermion loop in the $b^\mu$ background was studied.  The
results of these studies have been controversial, since they
depended on the type of regularization adopted during
calculations. The final conclusion can only be drawn after certain
additional physical assumptions have been made. New interesting
results at finite temperature have also been obtained, --- for
instance, the Chern--Simons term vanishes in the very high
temperature limit \cite{15}. The applications  of  the SME
predictions to astrophysics and cosmology are also of interest
\cite{1c} (see, also \cite{14}, where, in particular, energy
splitting between electrons of different helicities and decay of
very high energy electrons into lower energy electrons and
positrons have been considered). It should also be mentioned that
this type of axial-vector interaction arises in the study of the
coherent interaction of neutrino with dense matter in the
framework of the Standard Model (see, e.g., \cite{16}).

The parameter $b^\mu$ as a 4-vector  may be time-like or
space-like. In what follows, we shall consider the first
possibility. The experimental bounds for the space components of
$b^\mu$ are rather stringent, i.e., $|{\bf b}| <
10^{-18}$--$10^{-20}$eV, or even stronger, depending on the type
of the experiment (for more details of the experimental situation,
see \cite{exp,new}, and references therein). Interesting enough is
the fact that no stringent experimental limitations for the time
component $b^{0}$ (at least for an electron, and this case is just
of interest for us) have yet been obtained. One may argue (see,
e.g., \cite{14}) that, since the precision for measurements of the
electron mass is about $10^{-8}$, the upper bound on the time
component may be $|b_0|<10^{-2}$eV.

In the present work we investigate the possibility of
electron-positron pair production by a photon and  radiation of a
photon by electrons and positrons induced by the Lorentz breaking
background $b^{0}$. We assume that the photon dispersion law
remains unchanged,  and hence we neglect Cherenkov radiation. Our
calculations are made in the framework of the Furry picture
\cite{F51}, i.e., we consider the  Lorentz breaking axial vector
background as a kind of an external field and take it into
consideration exactly, i.e., without making any expansion in
powers of the parameter $b^{0}.$
 This means that creation and annihilation operators are exact
  solutions of the Dirac equation in the external or condensate field
  and form the basis for the perturbation theory description of
  interaction with the radiation field.  This technique is
appropriate when external or condensate fields are strong and
particle energies are high. Moreover, since the dispersion law for
charged particles in the external or condensate field changes as
compared with the zero field case, new channels of reactions may
become open. The calculations with this basis are, however, quite
complicated. Nevertheless, it is with the use of this technique
that various processes feasible in the astrophysics, such as
one-photon electron-positron pair production, photon splitting,
beta-decay in the strong magnetic field of pulsars,  and a number
of others were investigated (see, e.g.,
\cite{klepikov,adler,ritus,shabad,L16,ZZ}; for the recent reviews
of such calculations, see \cite{kniga,BVZ97}).

\section{The Model}
Consider fermions interacting with an electromagnetic field
$A^{\mu}(x)$ and with a constant condensate field $b^{\mu}$. The
Lagrangian density of the model \cite{2} is as follows:
\begin{equation}
 {\cal L}={\cal L}_{\rm{em}}+{\cal L}_{\rm{Dir}},
 \label{l}
\end{equation}
where
\begin{equation}{\cal L}_{\rm{em}}=-\frac{1}{4}F_{\mu\nu}F^{\mu\nu}
\end{equation}
is the electromagnetic field Lagrangian and
\begin{equation}
{\cal L}_{\rm{Dir}}=\overline{\psi}(i\gamma^{\mu}
\partial_{\mu}+e\gamma^{\mu}A_{\mu}-m-b_{\mu}\gamma^{\mu}\gamma_5)\psi
 \label{L}\end{equation}
is the Lagrangian of the Dirac field.

In order to calculate the rate of pair photo-production and the rate
of radiative transition of an electron induced
by the Lorentz breaking background in the framework of the Furry
picture, the ``extended'' Dirac equation
that takes into account interaction with the axial vector term should
be solved. As it follows from the
SME Lagrangian (\ref{L}), the Dirac equation has the form
\begin{equation}\label{1}
  \left(i
  {\gamma^{\mu}{\partial}_{\mu}}-\gamma^{\mu}{b_{\mu}}\gamma^{5}-m\right)
  \psi=0.
\end{equation}

The  canonical momentum operator $i{{\partial}^{\mu}}$  commutes
with the  operator of equation (\ref {1}). It can be shown (see
\cite{L05}) that the eigenvalues of the operator $i\partial^{\mu}$
are as follows:
\begin{equation}\label{a20} P^{\mu}= q^{\mu}\left(1+\frac{\zeta
b^{2}}{\sqrt{(bq)^{2}-b^{2}m^{2}}}\right)
+{b^{\mu}}\left(1-\frac{\zeta
(bq)}{\sqrt{(bq)^{2}-b^{2}m^{2}}}\right),
\end{equation}
where $q^\mu$ is a constant 4-vector, such as $ q^{2}= m^{2},$ and
$\zeta = \pm 1$. The dispersion law that follows from eq.
(\ref{a20}) has the form
\begin{equation}\label{b20}
P^{2}= m^{2}+2(Pb)-2{b^{2}}- 2\zeta
\sqrt{\left((Pb)-{b^{2}}\right)^{2}-b^{2}m^{2}}.
\end{equation}

Consider
now the most interesting case of a time-like vector
 $b^\mu$ and choose it in the form $b^{\mu}=\{b,0,0,0\}.$ In this
 case the
orthonormalized system of solutions of equation (\ref{1}) can be
written as follows:
\begin{equation}\label{0a21}
\psi(x)= \frac{\left|\Delta_{q \zeta}\right|}{\sqrt{2q^{0}}}
\,e^{-iq^{0}x^{0}} e^{i{\bf{qx}}\Delta_{q \zeta} }
(\gamma^{\mu}{q_{\mu}}+m)(1-\zeta\gamma^{5}\gamma_{\nu}{S}^{\nu}_{tp}
)\psi^0,
\end{equation}
\noindent where $\psi^0$ is a constant bispinor, $\Delta_{q \zeta} =
1+\zeta b/|{\bf{q}}|,$ and
\begin{equation}\label{0021}
S_{tp}^{\mu}=\frac{1}{m}\left\{|{\bf q}|,
 q^{0}{\bf q}/|{\bf q}|\right\}.
\end{equation}

It is easily seen that the 4-vector $q^{\mu}=\{q^0, {\bf q}\}$
plays the role of the kinetic momentum of the particle with the
energy $\varepsilon = q^0$ given by the same dispersion relation
as in the case of a free particle in the Lorentz invariant theory
\begin{equation}\label{003}
    \varepsilon = \sqrt{{\bf q}^{2}+m^{2}}.
\end{equation}

It is clear that in this case the relation between the canonical
momentum $\bf{P}$ and the kinetic momentum $\bf{q}$ is determined
by the formula
\begin{equation}\label{03}
    {\bf P} = {\bf q}\Delta_{q \zeta},
\end{equation}
and consequently,  the dispersion law can be rewritten in the form
\begin{equation}\label{22}
\varepsilon = \sqrt{\left(\Delta |{\bf{P}}|- \zeta
b\right)^{2}+m^{2}},
\end{equation}
\noindent where $\Delta = {\mathrm{sign}\left(\Delta_{q \zeta}
\right)}$ is the sign factor. It is therefore clear that
\begin{equation}\label{velocity}
{\bf v}=\frac{\partial \varepsilon}{\partial {\bf{P}}}=\frac{\bf{q}}{q^{0}}
\end{equation}
\noindent is  the particle group velocity.

The relation (\ref{22}) differs from those used in papers
\cite{16} by the sign factor $\Delta.$ This is due to the fact
that, in those papers, the canonical momentum  ${\bf{P}}$ and not
the kinetic momentum ${\bf q}$ was used as the particle quantum
number, and $\zeta$ was the projection of the particle spin on the
canonical momentum. It should be emphasized however, that the
particle kinetic momentum components, related to the group
4-velocity $u^{\mu}$ by the relation $q^{\mu} = mu^{\mu},$ $
q^{2}= m^{2},$ and not the canonical momentum components, are
suitable to play the role of the particle quantum numbers.
Moreover, we choose the helicity of the particle  as the spin
quantum number $\zeta$. It is well known that for a particle in an
external field, the projection of its spin on the direction of its
kinetic momentum is defined as its helicity \cite{ST,BG,T}. In our
problem the directions of canonical and kinetic momenta are
different, and hence, the projection of particle spin on the
canonical momentum does not coincide with its helicity. Thus, we
believe that the particle kinetic momentum, related to its
group velocity (\ref{velocity}), and  its helicity are the
particle physical variables  that can be considered as its
observables. This justified our decision that they should be taken
as the particle quantum numbers.

\section{Pair production}

Consider the electron-positron pair photo-production process. The
probability of pair production by a polarized photon is defined by
the relation
\begin{equation}\label{ar1}
\begin{array}{c}
  \displaystyle W=\frac{1}{2k^{0}}\!\int\!\! d^{4}x\,d^{4}y\!\int
  \frac{d^{4}p\,d^{4}q}{(2\pi)^{6}}\,
  \delta(p^{2}\!-m^{2})\delta(q^{2}\!-m^{2})
  \\[8pt]
\displaystyle \times {\mathrm{Sp}}
\big\{\gamma_{\mu}(x)\varrho_{\bar{e}}(x,y;p,\zeta_{p})
\gamma_{\nu}(y)\varrho_{e}(y,x;q,\zeta_{q})\big\}
\varrho^{\mu\nu}_{ph}(y,x;k).
\end{array}
\end{equation}
\noindent \noindent Here,
$\varrho_{e}(y,x;q,\zeta_{q}),\varrho_{\bar{e}}(x,y;p,\zeta_{p})$
are the electron and positron density matrices respectively,
$\,\varrho^{\mu\nu}_{ph}(y,x;k)$ is the density matrix of the
initial photon with 4-momentum $k^{\mu}$. The density matrices of
longitudinally polarized electron with the 4-momentum $q^{\mu}$
and helicity $\zeta_q$ and positron with 4-momentum $p^{\mu}$ and
helicity $\zeta_p$ in the Lorentz breaking background constructed
with the use of solutions (\ref{0a21}) have the form
\begin{equation}\label{arx4}
\begin{array}{c}
\displaystyle \varrho_{\bar{e}}(x,y;p,\zeta_{p})=
\frac{1}{2}\Delta_{p\zeta_{p}}^{2}(\gamma^{\mu}{p_{\mu}}- m)
(1-\zeta_{p}\gamma^{5}\gamma_{\nu}{S}^{\nu}_{tp}(p)) \displaystyle
e^{i(x^{0}-y^{0})p^{0}-i({\bf x}-{\bf y}) {\bf
p}\Delta_{p\zeta_{p}}},\\[8pt]
\displaystyle \varrho_{{e}}(y,x;q,\zeta_{q})=\frac{1}{2}
\Delta_{q\zeta_{q}}^{2}(\gamma^{\mu}{q_{\mu}}+m)
(1-\zeta_{q}\gamma^{5}\gamma_{\nu}{S}^{\nu}_{ tp}(q))
\displaystyle  e^{i(x^{0}-y^{0})q^{0}-i({\bf x}-{\bf y}) {\bf
q}\Delta_{q\zeta_{q}}}.
\end{array}
\end{equation}

Upon integrating with respect to coordi\-nates, we obtain the
following expression for the transition rate under
investi\-ga\-tion:
\begin{equation}\label{arxx4}
\begin{array}{c}
\displaystyle W= \frac{e^{2}}{k^{0}}\!\!\int
  \frac{d^{4}p\,d^{4}q}{2\pi}\,
  \delta(p^{2}\!-m^{2})\delta(q^{2}\!-m^{2})\delta(k^{0}-p^{0}-q^{0})
 \delta^{3}
  ({\bf
  k} -{\bf p}\Delta_{p\zeta_{p}}-{\bf q}\Delta_{q\zeta_{q}})T(p,q).
\end{array}
\end{equation}
\noindent Here
\begin{equation}\label{arxxx4}
T(p,q)=\displaystyle\frac{1}{4}{\mathrm{Sp}}\left\{
\gamma^{\alpha}{e_{\alpha}}(\gamma^{\mu}{p}_{\mu}-
m)(1-\zeta_{p}\gamma^{5}\gamma_{\nu}{S}^{\nu}_{tp}(p))
{\gamma^{\beta} {e}^{*}_{\beta}}(\gamma^{\rho}{q_{\rho}}+m)
(1-\zeta_{q}\gamma^{5}\gamma_{\lambda}{S}^{\lambda}_{tp}(q))
\right\},
\end{equation}
where $e^{\mu}=\left\{0,{\bf a}_{1}+ig{\bf
a}_{2}\right\}/{\sqrt{2}}$, with $g=\pm 1$, are unit vectors of circular
polarization.

Instead of the electron, positron and photon energies
$\varepsilon_{{e}}\equiv q^0,\,\varepsilon_{\bar{e}}\equiv p^0,
\varepsilon_{\gamma}\equiv k^0$ respectively, and polarizations
$\zeta_q,\,\zeta_p,\, g,$ and the effective
potential $b$, it is convenient to introduce the dimensionless quantities
\begin{equation}\label{q10}
\begin{array}{llll}
\eta = k^{0}/2m,\; & \eta+y=p^{0}/m,\; &
 \eta-y= q^{0}/m,&{}\\[4pt]
\bar{g}={g}\,{\mathrm{sign}}(b),&
\bar{\zeta}_{\bar{e}}={\zeta}_{p}\,{\mathrm{sign}}(b),&
\bar{\zeta}_{e}={\zeta}_{q}\,{\mathrm{sign}}(b),  & d=|b|/m.
\end{array}
\end{equation}
Then the results of integration in (\ref{arxx4}) can be expressed in the form
\begin{equation}\label{q16}
\begin{array}{c}
\displaystyle W_{\bar{g}\bar{\zeta}_{\bar{e}}\bar{\zeta}_{e}} =
\frac{e^{2}m}{32\eta^{2}}\int\!\!
  \frac{dy}{\sqrt{(\eta+y)^{2}-1}\sqrt{(\eta-y)^{2}-1}}\\[12pt]
 \displaystyle\times \left[
  \bar{\zeta}_{\bar{e}}\sqrt{(\eta+y)^{2}-1}+\bar{\zeta}_{e}
  \sqrt{(\eta-y)^{2}-1} +2d +
  2 \bar{g}\eta \right]^{2}.
\end{array}
\end{equation}
\noindent The integration limits  in the above formula depend on
the values of parameter $d.$ In the realistic limit of small $d\ll
1$, for the case $\bar{\zeta}_{\bar{e}}=1, \bar{\zeta}_{{e}}=1,$
we have \cite{x}
\begin{equation}\label{q011}
\begin{array}{ll}
y\in\varnothing & \quad \eta \in [1,\eta_{1}), \\
y\in [-y_{0},y_{0}] & \quad \eta \in [\eta_{1},\infty),
\end{array}
\end{equation}
and for other values of $\bar{\zeta}_{\bar{e}}$ and  $\bar{\zeta}_{e}$
\begin{equation}\label{q11}
\begin{array}{ll}
y\in\varnothing & \quad \eta \in [1,\infty),
\end{array}
\end{equation}
where
\begin{equation}\label{q13}
\eta_{1}= \frac{1+d^{2}}{2d},\qquad y_{0}=\frac{|\eta
-d|\sqrt{\eta -\eta_{1}}}{\sqrt{\eta -d/2}}.
\end{equation}

In other words, there exists a threshold $\eta_1$, depending on
the
parameter $b$, such that only for $\eta \geqslant\eta_1$ the
process can take place. This is natural for the $e^+ e^- $ pair
production process in the background field. Moreover, as it is
seen from (\ref{q10}), (\ref{q011}), (\ref{q11}), only those
electrons and positrons can be produced, whose helicities are
equal and whose sign coincide with the sign of the effective
potential  $b$.

Then, upon integrating over $y$, one obtains the transition rate
\begin{equation}\label{j0}
    W_{\bar{g}\bar{\zeta}_{\bar{e}}\bar{\zeta}_{e}}=
    \frac{e^{2}m}{32\eta^{2}}(1+\bar{\zeta}_{\bar{e}})
    (1+\bar{\zeta}_{{e}}) \Theta (\eta-\eta_{1})\, J.
\end{equation}
\noindent Here  $\Theta (\eta-\eta_{1})$ is the Heaviside step
    function and
\begin{equation}\label{j1}
\begin{array}{c}
   \displaystyle J = \left[2(\eta +\bar{g}d) +d^{2}/\eta
   \right]F(\chi,s)
   - \eta  E(\chi,s)\\[8pt]+ \displaystyle y_{0}\left[1+ \frac
   {\sqrt{\left((\eta-y_{0})^{2}-1\right)}\sqrt{\left((\eta+y_{0})^{2}-1\right)}}
   {{y_{0}^{2}+\eta^{2}-1}}\, \right] \\[12pt]
\displaystyle  +2\left(d+\bar{g}\eta\right)\ln
    \frac{\sqrt{\left((\eta+y_{0})^{2}-1\right)}+\eta +y_{0}}
    {\sqrt{\left((\eta-y_{0})^{2}-1\right)}+\eta -y_{0}},
\end{array}
\end{equation}
\noindent where $F(\chi,s),E(\chi,s)$ are the elliptic integrals
\cite{BE} of the arguments
\begin{equation}\label{j2}
    \chi = \arcsin \frac{2y_{0}\eta }{y_{0}^{2}+\eta^{2}-1},\qquad
    s=\sqrt{1-\eta^{-2}}\,.
\end{equation}

In the limit of small $d \ll 1, $ the following
expression can be obtained from (\ref{j0})
\begin{equation}\label{j5}
 W_{\bar{g}\bar{\zeta}_{\bar{e}}\bar{\zeta}_{e}}\approx
    \frac{e^{2}m}{16\eta}\, \ln \frac{1+ \sqrt{1-\eta_{1}/\eta}}
    {1- \sqrt{1-\eta_{1}/\eta}}(1+\bar{\zeta}_{\bar{e}})
    (1+\bar{\zeta}_{{e}})(1+
    \bar{g}).
\end{equation}
\noindent Near the reaction threshold
$\eta =
    \eta_{1}\approx ({2d})^{-1}$, the transition rate is described
by the formula
\begin{equation}\label{j3}
 W_{\bar{g}\bar{\zeta}_{\bar{e}}\bar{\zeta}_{e}}\approx
    \frac{e^{2}m}{8\eta}\, \sqrt{1-\eta_{1}/\eta}(1+\bar{\zeta}_{\bar{e}})
    (1+\bar{\zeta}_{{e}})(1+
    \bar{g}),
\end{equation}
\noindent and high above the threshold, with $d\eta \gg 1$, by the formula
\begin{equation}\label{j4}
 W_{\bar{g}\bar{\zeta}_{\bar{e}}\bar{\zeta}_{e}}\approx
    \frac{e^{2}m}{16\eta}\, \ln (8d\eta)\,(1+\bar{\zeta}_{\bar{e}})
    (1+\bar{\zeta}_{{e}})(1+
    \bar{g}).
\end{equation}

\begin{figure}[h!!!]
\begin{center}
\includegraphics[width=0.45\textwidth]{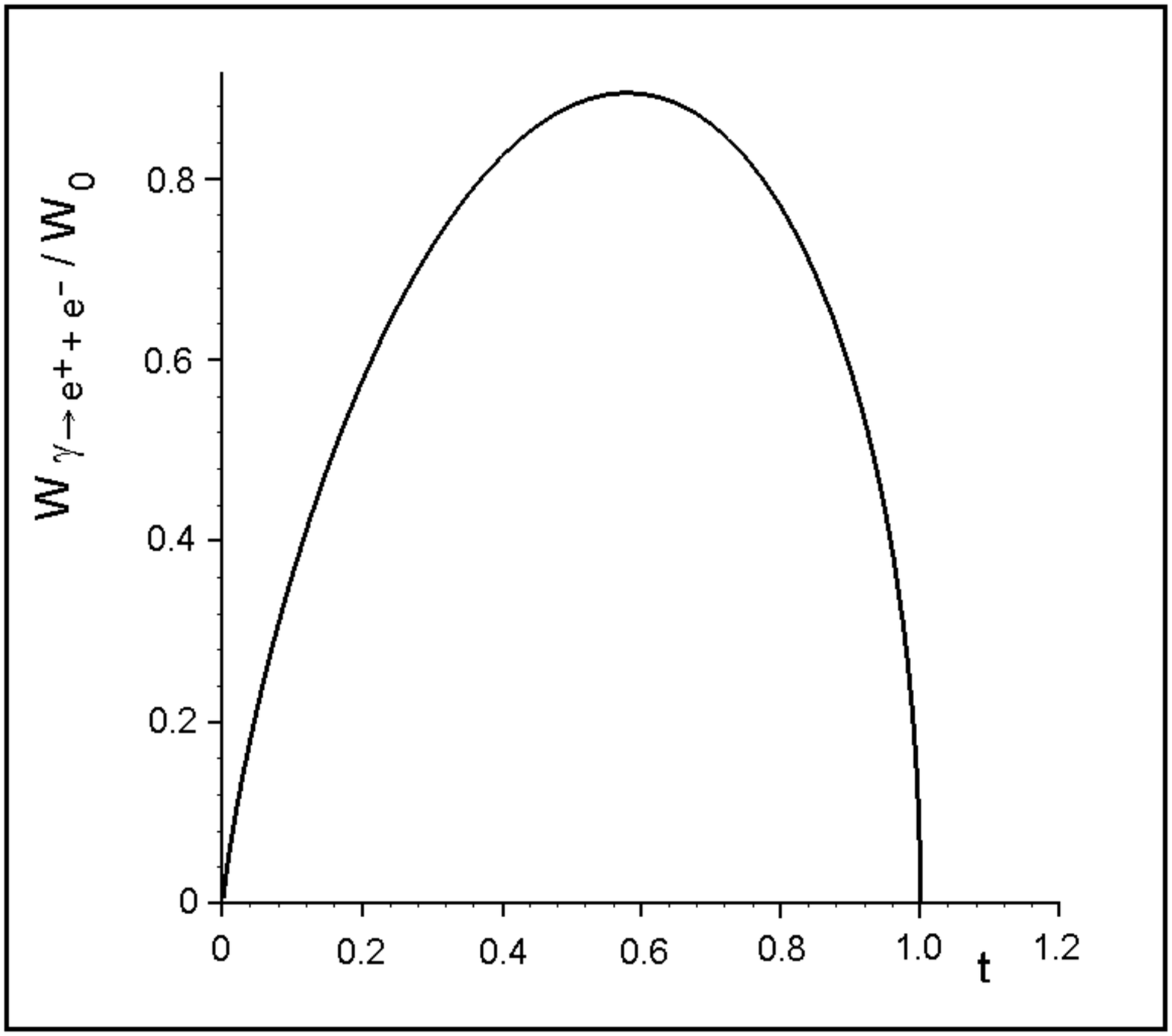}
\caption{The rate of electron-positron pair production.
 $
W_{0}=e^{2}m/2\eta_{1} \approx 10^{13}\left(|b|/{\mathrm{1
eV}}\right ){\mathrm{sec}}^{-1}$; $t= \eta_{1}/\eta\approx
2.5\cdot 10^{11}({\mathrm{1 eV}})^{2}/|b|\varepsilon_{\gamma}.$}
\label{pair1}
\end{center}
\end{figure}

It follows from the last formulas (\ref{j3}) and (\ref{j4}) that,
in the realistic case of small $d\ll 1$, only the photons with the
helicity $g$, whose sign coincides with the sign of the effective
potential, i.e., $\bar{g}=1$, can effectively produce pairs.
This is why the threshold singularity in the
transition rate is of the square root form, which is
characteristic for the processes allowed by the angular momentum
selection rule. For photons with the opposite helicity sign, the
pair production is suppressed. The rate of the process as a
function of the inverse photon  energy is depicted in FIG.
\ref{pair1}.

\section{Photon emission}

Consider now  the cross-channel, i.e., photon emission by a
charged particle in the Lorentz violating background. The authors
of \cite{1e} proposed to call this process ``helicity decay''. In
fact, this is an example of the  more general sort of radiation
previously called ``spin light'' \cite{BTB95}, which is just
radiation of an intrinsic magnetic moment of an electron
associated with its spin. This sort of radiation has been studied
for the last years in a number of papers. In the case of the
``synchrotron radiation'', i.e., radiation of a relativistic
charged particle in an external magnetic field,  its dependence on
the electron spin orientation  was studied both theoretically
\cite{theor} and experimentally \cite{exper}. As a result of these
studies, it has become clear that the synchrotron radiation can be
considered as consisting of two parts: one is radiation of the
electron charge itself and the other is just the radiation of an
intrinsic magnetic moment of an electron, and it is just what they
called the ``spin light''. This term was also used in \cite{L49}
for the description of photon radiation by a neutrino in dense
media. It should be emphasized that the mechanism of this last
process (see \cite{L05}) is similar to that of the process under
investigation in the present article. As it will be seen from the
formulas to follow, radiative transitions can take place both with
and without the spin flip. This is why we prefer to use the term
``spin light'', and not the ``helicity decay'' in our paper.

First of all we point out that the formulas we obtain in what follows are
valid  for  both an electron and  a positron. This is due to the
fact that the sign in front of the $\gamma^{5}$ matrix in equation
(\ref{1}) remains invariant under the charge conjugation
operation.

The rate of transition of an electron from the initial state with
4-momentum $p^{\mu}$ and helicity $\zeta_{p}$ to the final state
$q^{\mu},\,\zeta_{q}$ with emission of a photon with circular
polarization $g$ can be written in the form
\begin{equation}\label{j7}
\displaystyle W_{\bar{g}\bar{\zeta}_{i}\bar{\zeta}_{f}} =
\frac{e^{2}m}{8\gamma(\gamma^{2}-1)}\int
  \frac{dx}{\sqrt{x^{2}-1}}
  \displaystyle \left[\bar{\zeta}_{i}{\sqrt{\gamma^{2}-1}}+\bar{\zeta}_{f}
  {\sqrt{x^{2}-1}}+
     2d+\bar{g}(\gamma-x)\right]^{2},
\end{equation}
where  instead of the initial and final electron energies
$\varepsilon_{i}\equiv p^0,\;\varepsilon_{f}\equiv q^0$
respectively and particle polarizations $\zeta_p,\,\zeta_q$
we introduced, in addition to (\ref{q10}), the following dimensionless
variables:
\begin{equation}\label{j6}
\gamma=p^{0}/m,\quad x=q^{0}/m,\quad
\bar{\zeta}_{i}={\zeta}_{p}\,{\mathrm{sign}}(b),\quad
\bar{\zeta}_{f}={\zeta}_{q}\,{\mathrm{sign}}(b).
\end{equation}
\noindent The integration limits in the formula (\ref{j7}) are
\begin{equation}\label{q0111}
\begin{array}{ll}
x\in \varnothing & \quad \gamma \in [1,\infty),
\end{array}
\end{equation}
\noindent if $\bar{\zeta}_{i}=1,$ and
\begin{equation}\label{q111}
\begin{array}{ll}
x\in \varnothing & \quad \gamma \in [1,\gamma_{0}),\\
x\in[\omega_{1},\omega_{2}] & \quad \gamma \in
[\gamma_{0},\gamma_{1}),\\  x\in[1,\omega_{2}] & \quad \gamma \in
[\gamma_{1},\gamma_{2}),\\  x\in \varnothing &
\quad \gamma \in [\gamma_{2},\infty),\\
\end{array}
\end{equation}
\noindent if $\bar{\zeta}_{i}=-1,\bar{\zeta}_{f}= -1,$ and
\begin{equation}\label{q12}
\begin{array}{ll}
x\in \varnothing &
\quad \gamma \in [1,\gamma_{1}),\\
x\in[1,\omega_{1}] &
\quad \gamma \in [\gamma_{1},\gamma_{2}),\\
x\in[\omega_{2},\omega_{1}] &
\quad \gamma \in [\gamma_{2},\infty),\\
\end{array}
\end{equation}
\noindent if $\bar{\zeta}_{i}=-1,\bar{\zeta}_{f}= 1.$

\noindent Here
\begin{equation}\label{q131}
\omega_{1}= \frac{1}{2}\left(z_{+1} + z_{+1}^{-1}\right),\qquad
\omega_{2}= \frac{1}{2}\left(z_{-1} + z_{-1}^{-1}\right),
\end{equation}
where
\begin{equation}\label{q14}
\begin{array}{l}
z_{ \pm 1}=\displaystyle \gamma \pm
\left(\sqrt{\gamma^{2}-1}-2d\right),
\end{array}
\end{equation}
and
\begin{equation}\label{q15}
\begin{array}{ll}
\gamma_{0}=\displaystyle \sqrt{1+d^{2}}, & {}\\[8pt]
\displaystyle \gamma_{1}=\frac{1}{2}\left\{\left(1+2d\right)
 +\left(1+2d\right)^{-1}\right\},& {}\\[12pt]
 \displaystyle \gamma_{2}=\frac{1}{2}\left\{\left(1-2d\right)
 +\left(1-2d\right)^{-1}\right\}.& {}
\end{array}
\end{equation}

\noindent As it is seen from the  above formulas, radiative
transitions can take place both with and without the spin flip.

The integration is carried out elementary and we obtain
\begin{equation}\label{q18}
\begin{array}{c}
\displaystyle
W_{\bar{g}\bar{\zeta}_{i}\bar{\zeta}_{f}}=\frac{e^{2}m}{16}
\Big\{(1+ \bar{\zeta}_{f})\left[Z(z_{+1},1)
  \Theta (\gamma -\gamma_{1})
  +Z(z_{-1},-1)\Theta(\gamma -\gamma_{2})\right]\\[6pt]
\displaystyle +(1-\bar{\zeta}_{f})\left[Z(z_{+1},1)
  \Theta(\gamma_{1} -\gamma)
  +Z(z_{-1},-1)\Theta(\gamma_{2} -\gamma)\right]
  \Theta(\gamma -\gamma_{0})\Big\}(1-\bar{\zeta}_{i}).
\end{array}
\end{equation}
\noindent Here
\begin{equation}\label{q17}
\begin{array}{c}
\displaystyle
Z(z,\bar{\zeta}_{f})=\frac{1}{8\gamma(\gamma^{2}-1)}\Big\{4z_{-\bar{g}}^{2}\ln
z+ \Big[\left(z^{2}-z^{-2}\right)-\bar{g}\bar{\zeta}_{f}
\left(z-z^{-1}\right)^{2}\Big] \\[10pt]
-4z_{-\bar{g}} \left[\left(z-z^{-1}\right)-\bar{g}\bar{\zeta}_{f}
\left(z+z^{-1}-2\right)\right]\Big\}.
\end{array}
\end{equation}
\noindent After summation  over polarizations of the final
particle, the transition rate becomes
\begin{equation}\label{q19}
W_{\bar{\zeta}_{f}=1}+W_{\bar{\zeta}_{f}=-1}=
\frac{{e^{2}m}}{8}(1-\bar{\zeta}_{i})\Big\{Z(z_{+1},1)
+Z(z_{-1},-1)\Big\} \Theta(\gamma -\gamma_{0}).
\end{equation}
\noindent If $d\gamma \ll 1,$  expression (\ref{q18}) leads to the
formula
\begin{equation}\label{q20}
W_{\bar{g}\bar{\zeta}_{i}\bar{\zeta}_{f}}\approx\frac{2e^{2}m
d^{3}}{3}
\gamma(\gamma^{2}-1)^{1/2}(1-\bar{g}\beta)^{2}(1-\bar{\zeta}_{i})(1+\bar{\zeta}_{f}),
\end{equation}
\noindent where $\beta = \sqrt{\gamma^{2}-1}/\gamma$ is the
initial  particle  velocity.

In the relativistic limit $(\gamma
\gg 1),$ the transition rate is transformed to the expression
\begin{equation}\label{q21}
W_{\bar{g}\bar{\zeta}_{i}\bar{\zeta}_{f}}\approx\frac{e^{2} m
}{16\gamma}\left[\ln(1+4d\gamma)-\frac{4d\gamma(1+6d\gamma)}
{(1+4d\gamma)^{2}}\right]
(1-\bar{\zeta}_{i})(1+\bar{\zeta}_{f})(1-\bar{g}),
\end{equation}
which is valid for $d \ll 1$.

\begin{figure}[h!!!]
\begin{center}
\includegraphics[width=0.45\textwidth]{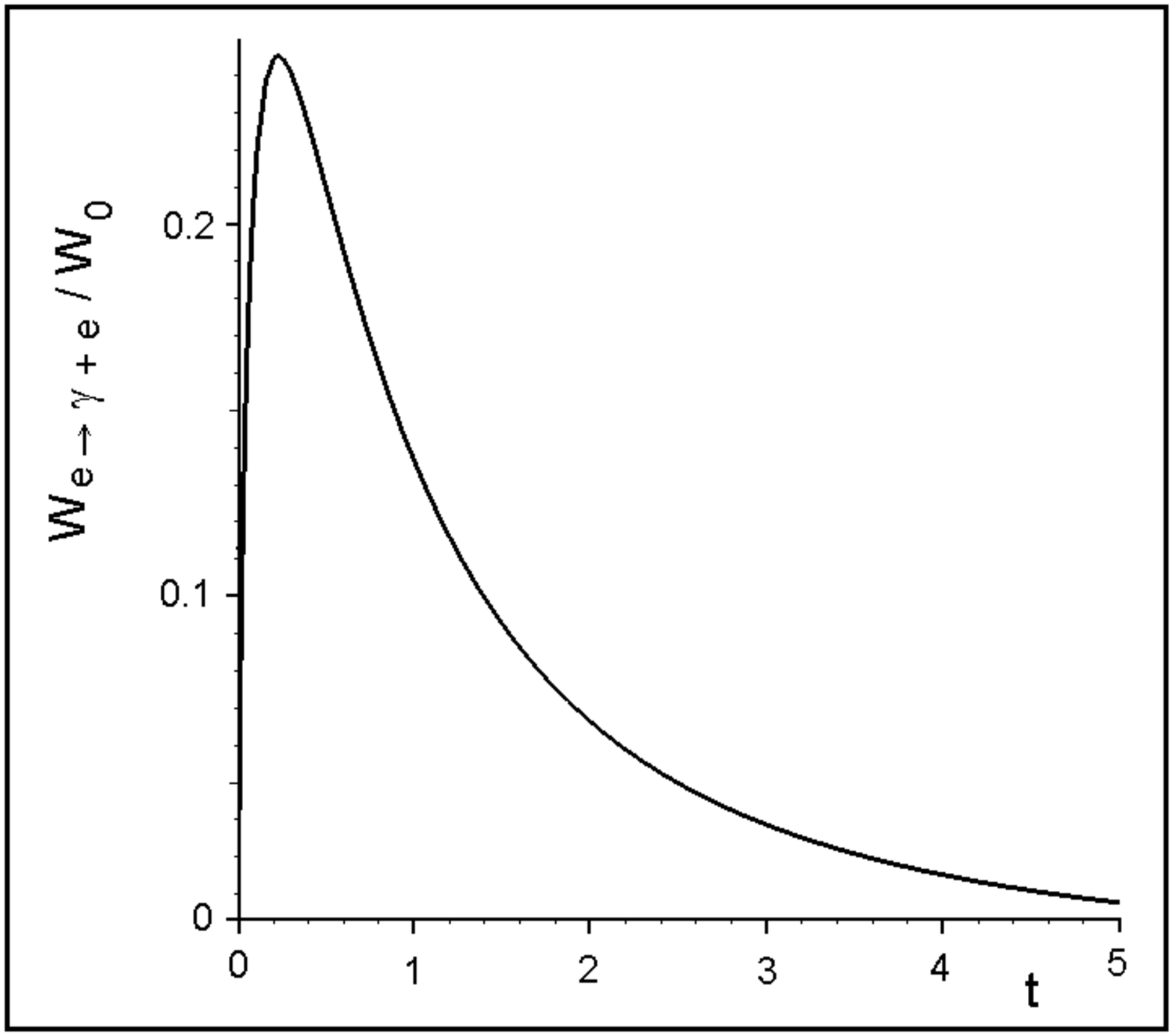}
\caption{{The rate of the radiative transition.$ W_{0}=2e^{2}md
\approx 2\cdot 10^{13}\left(|b|/{\mathrm{1 eV}}\right
){\mathrm{sec}}^{-1}$; $t= (4d\gamma)^{-1}\approx 0.6\cdot
10^{11}({\mathrm{1 eV}})^{2}/|b|\varepsilon_{i}.$}}\label{light1}
\end{center}
\end{figure}

Thus, we see that at high energies $\gamma \gg 1$ charged particles  radiate
primarily photons with the helicity sign opposite to the sign of
the effective potential $b$ (${\bar g}=-1$). Moreover, the radiation process is
accompanied by the particle helicity flip. The rate of the process
as a function of the electron inverse energy is depicted in
FIG.\ref{light1}.

Let us consider now the radiation power of spin light. If we
introduce the function
\begin{equation}\label{q22}
  \tilde{Z}(z,\bar{\zeta}_{f}) = \gamma Z(z,\bar{\zeta}_{f})
   - Y(z,\bar{\zeta}_{f}),
\end{equation}
\noindent where
\begin{equation}\label{q23}
\begin{array}{c}
\displaystyle
Y(z,\bar{\zeta}_{f})=\frac{1}{24\gamma(\gamma^{2}-1)}\Big\{-12z_{-\bar{g}}\ln
z - 3z_{-\bar{g}} \Big[\left(z^{2}-z^{-2}\right)
-\bar{g}\bar{\zeta}_{f} \left(z-z^{-1}\right)^{2}\Big]\\[8pt]
+\displaystyle \Big[(z-z^{-1})^{3}-\bar{g}\bar{\zeta}_{f}
\left((z+z^{-1})^{3}-8\right)\Big]+6(z-z^{-1})(z_{-\bar{g}}^{2}+1)\Big\},
\end{array}
\end{equation}
\noindent then the formula for the total radiation power can be
obtained from (\ref{q18}), (\ref{q19}) by  substitution
${Z}(z,\bar{\zeta}_{f}) \rightarrow \tilde{Z}(z,\bar{\zeta}_{f}).$
It can be verified that if  $d\gamma \ll 1,$  the radiation power
becomes
\begin{equation}\label{q24}
\begin{array}{c}
\displaystyle
I_{\bar{g}\bar{\zeta}_{i}\bar{\zeta}_{f}}\approx\frac{2e^{2}m^{2}d^{4}}{3}
(\gamma^{2}-1)(1-\beta\bar{g})(3\gamma^{2}(1-\beta\bar{g})-1)
(1-\bar{\zeta}_{i})(1+\bar{\zeta}_{f}).
\end{array}
\end{equation}
\noindent  In the relativistic limit, the radiation power is equal
to
\begin{equation}\label{q25}
\begin{array}{c}
\displaystyle
I_{\bar{g}\bar{\zeta}_{i}\bar{\zeta}_{f}}\approx\frac{e^{2} m
^{2}}{16}\left[\ln(1+4d\gamma)-\frac{2d\gamma(6+15(4d\gamma)+11(4d\gamma)^{2})}
{3(1+4d\gamma)^{3}}\right]
(1-\bar{\zeta}_{i})(1+\bar{\zeta}_{f})(1-\bar{g}).
\end{array}
\end{equation}
It should be first of all emphasized that as follows from our
result for the radiation power (\ref{q25}), photons emitted  by
relativistic particles ($\gamma\gg 1$) are completely circulary
polarized, $\bar {g}=-1$, and the helicity sign in this case is
opposite to the sign of the effective potential $b$. Moreover, the
total radiation power  increases logarithmically. It can be seen
from equations (\ref{q21}) and (\ref{q25}) that in the
ultra-relativistic limit ($d\gamma \gg 1$) the average energy of
emitted photons $\langle\varepsilon_{\gamma}\rangle$ is almost
equal to the electron initial energy $\varepsilon_{i}$
\begin{equation}\label{q26}
\langle\varepsilon_{\gamma}\rangle\approx \varepsilon_{i}\left[ 1
-\frac{1}{3(\ln(1+4d\gamma)- 3/2)}\right].
\end{equation}
\noindent It is therefore evident that the  particle looses a
rather significant amount of its  initial energy through
radiation.

\section{Discussion and conclusions}

The results obtained demonstrate the strong dependence of the
rates on the polarization properties
of the particles that participate and are produced in the
reactions we studied.  In general, this
amounts to the conclusion that, in the framework of the present
model, in the ultra-relativistic limit, the particles produced in the
reactions considered are preferably in the state
with determined polarization, i.e., photons have the sign of
polarization opposite to the sign of the effective potential,
while charged particles are preferably in the state with the
helicity coinciding with the sign of the effective potential. This
certainly is the consequence of the
particular choice of the model with Lorentz symmetry violating
term in the Dirac Lagrangian we adopted. Indeed, the model
assumes axial-vector interaction with the background field, and
this appears to be essential to the helicity selection rules. The
question arises, as to what may happen for other Lorentz
structures such as vector, tensor, and chiral structures such as
$(V \pm A)$ in the Lagrangian that also occur in a reasonable
standard model extension. In the standard model extension \cite{2}
(we restricted ourselves to its electrodynamic sector only),
interactions of particles with various condensate fields of the vector,
tensor and axial-vector nature are included in the Lagrangian, and the
resulting theory is renormalizable. As is known, rates of the
processes in such theories should decrease
with growing energy in the very high energy limit. Hence, the system
in the final state should have zero orbital moment. The results of the
present paper confirmed this conclusion for the particular choice of the
axial-vector interaction with the background field. One may expect
that with other forms of the Lorentz violating background,
only those transitions would dominate that are allowed by this
general selection rule.
At the same time, for different types of interaction, this selection
rule may lead to different correlations between the spin quantum numbers
of participating particles. In the case in question, there
is no specified direction in space and these correlations are
quite simple.  On the other hand, e.g., with the
tensor type condensate determined by the parameter  $H^{\mu\nu}$ in the
SME Lagrangian, the fermion spin states are described by the
transversal polarization rather than helicity. Therefore, for this
case, any conclusions concerning spin correlation  can only be made
upon detailed consideration of the problem.

Due to the above mentioned polarization properties of particles
produced in the reactions considered, the photon radiation process
together with the pair production can not form a cascade process.
Let the sign of the effective potential be positive (in the
opposite case, the arguments are evident). Then as a result of the
radiative transition, a photon with left polarization and an
electron with arbitrary polarization are produced. However, the
right-handed electron can not radiate at all, while the
left-handed recoil electron, as is easily verified, will have the
energy lower than the threshold value $\gamma = \sqrt{1+ d^{2}}$.
Moreover, the rate of the pair production by the radiated photon
is practically equal to zero, as the rate of pair production by
the left-handed photon is strongly, $(\sim \eta_{1}^{-2})$,
suppressed as compared to the rate of pair production by the
right-handed photon.  Hence, the cascade process of the form
$$
e\rightarrow e +\gamma \rightarrow 2e + \bar{e},
$$
predicted in \cite{14}, proves to be impossible in our model,
at least in the resonance channel. Emission of
radiation by charged particles produced in the
process
$$
\gamma \rightarrow e + \bar{e}
$$
is also impossible in our model.

It should be mentioned that the results of the present work  are
different from corresponding estimates of
papers \cite{1c}. This is due to the fact that our results were
obtained with the use of exact Dirac wave functions and the corresponding
dispersion law for the particle energy in the Lorentz
violating background. The authors of \cite{1c}, however, without specifying any
mechanism,  postulated a modified dispersion law for an electron
such that it became different from the vacuum one by an additional
term cubic in the electron momentum.
As a result,
the threshold singularity of the pair creation process in
\cite{1c} had the form $({1-\eta_{1}/\eta})^{3/2}$ (in the notations
of our paper), and this is
characteristic for the processes forbidden by the angular momentum
selection rules. At the same time, the probability of photon
emission by an electron in \cite{1c} increases with growing energy as
$ \gamma^{8}$.

As it was already mentioned in the Introduction, at present there
are no serious experimental limitations for the value of parameter
$b$ in the case of an electron obtained in the low energy region.
Assuming, for instance, that the limitations are the same as for
the neutron \cite{exp}, we may come to the conclusion that the
effects considered above may become significant only at energies
comparable to the Planck energy $M_{{\mathrm{P}}}\sim
10^{28}{\mathrm{eV}}.$ This means that they may be important only
in the study of processes that take place in the early Universe.
In our calculations this corresponds to putting $d\sim
m/M_{{\mathrm{P}}}.$ The actual limitations for the value of
parameter $b$ can, however, be given only by experiment. The
results of the present paper can provide some possible guides for
obtaining the limitations on the value of $b$ in the high energy
region. The threshold for the process of pair production is fairly
high and hence it can be observed only in the astrophysical
conditions, while the spin light can be observed already in
laboratory experiments with modern accelerators.

To be particular, let us consider the following illustrative
example. With consideration for the above mentioned restriction
$b< 10^{-2}$eV, we have for the modern machines $d\gamma \ll 1.$
Therefore, with the use of equations (\ref{q20}) and (\ref{q24}),
we obtain, up to coefficient of order unity, for the transition
probability and spin light power respectively
\begin{equation}\label{y}
    W_{\mathrm {SL}}\sim
    \frac{\alpha}{\hbar}\,mc^{2}d\,(4d\gamma)^{2},\quad  I_{\mathrm {SL}}\sim
    \frac{\alpha}{\hbar}\,{\varepsilon_{i}}^{2}d^{2}\,(4d\gamma)^{2},
\end{equation}
where $\alpha$ is the fine structure constant, and the Gaussian
units were used.

Moreover, the average energy of emitted photons is defined by the formula
\begin{equation}\label{yy}
   \langle\varepsilon_{\gamma}\rangle_{\mathrm {SL}} \sim \varepsilon_{i}d\gamma.
\end{equation}
Now, for the same quantities in the case of synchrotron radiation, we
have the following
\begin{equation}\label{z}
    W_{\mathrm {SR}}\sim
    \frac{\alpha}{\hbar}\,mc^{2}(H/H_{0}),\quad  I_{\mathrm {SR}}\sim
    \frac{\alpha}{\hbar}\,{\varepsilon_{i}}^{2}(H/H_{0})^{2},
\end{equation}
where $H$ is the magnetic field strength, and $H_{0}=m^{2}c^{3}/e\hbar
\approx 4.41\cdot 10^{13}$G is the so called ``critical'', or
``Schwinger'' field (see, e.g., \cite{ST}). The average energy of
the synchrotron radiation photon is estimated as
\begin{equation}\label{zz}
   \langle\varepsilon_{\gamma}\rangle_{\mathrm {SR}} \sim \varepsilon_{i}
   (H/H_{0})\gamma.
\end{equation}
It is clear that in our problem, the parameter $b$ plays the same role
as the parameter $H/H_{0}$ in the synchrotron radiation
case. The presence of an extra small parameter $(4d\gamma)^{2}$ is due
to the different mechanism of radiation in our case: a photon is
emitted not by the electron charge but by its magnetic moment.
Therefore, we conclude that, in experiments with relativistic
electrons from modern accelerators, one may find certain limitations
on the value of the parameter $b$ by searching for possible hard
radiation from electrons in the straight parts of their trajectories,
where no synchrotron radiation should be expected.

\acknowledgments

The authors are grateful to D. Ebert and A. E. Shabad for fruitful
discussions.

\bigskip

This work was supported in part by the grant of President of
Russian Federation for leading scientific schools (Grant SS ---
2027.2003.2).


\begin{thebibliography}{99}
\bibitem{1}
 K. Hagiwara et al., Phys. Rev. D {\bf 66}, 010001 (2002).

\bibitem{2}
S. M. Carroll, G. B. Field, and R. Jackiw, Phys. Rev. D {\bf 41},
1231 (1990); D. Colladay and V. A. Kosteleck\'{y}, Phys. Rev. D
{\bf 55}, 6760 (1997), hep-ph/9703464; Phys. Rev. D {\bf 58},
116002 (1998), hep-ph/9809521; S. Coleman and S. L. Glashow, Phys.
Rev. D {\bf 59}, 116008 (1999), hep-ph/9812418.

\bibitem{new}
R. Bluhm, hep-ph/0506054.

\bibitem{teor} V. A. Kosteleck\'{y} and Ch. Lane, J.Math.Phys. {\bf 40}
6245 (1999), hep-ph/9909542; Phys.Rev. D {\bf 60},  116010 (1999),
hep-ph/9908504; R. Bluhm et. al, Phys.Rev. D {\bf 68}, 125008
(2003), hep-ph/0306190.

\bibitem{exp} D. F. Phillips  et al., Phys. Rev. D {\bf 63}, 111101 (2001);
M. A. Humphrey et al., Phys. Rev. A {\bf 68}, 063807 (2003),
physics/0103068; M. A. Humphrey, D. F. Phillips, and R. L.
Walsworth, Phys. Rev. A {\bf 62}, 063405 (2000); D. Bear et al.,
Phys. Rev. Lett. {\bf 85}, 5038 (2000); L.-S. Hou, W.-T. Ni, and
Y.-C. M. Li, Phys. Rev. Lett. {\bf 90}, 201101 (2003); R. Bluhm
and V. A. Kosteleck\'{y}, Phys. Rev. Lett. {\bf 84}, 1381 (2000);
F. Can\`{e} et al., Phys. Rev. Lett. {\bf 93}, 230801 (2004),
physics/0309070; P. Wolf et al., hep-ph/0509329.

\bibitem{1c}
T. Jacobson, S. Liberati, and D. Mattingly, Ann. Phys. (N.Y.) {\bf
321}, 150 (2006), astro-ph/0505267; D. Mattingly, Living Rev.
Relativity {\bf 8}, 5 (2005), gr-qc/0502097.

\bibitem{6}
I. L. Shapiro, Phys. Rept. {\bf 357}, 113 (2002), hep-th/0103093.

\bibitem{7}
G. E. Volovik, Pis'ma Zh. Eksp. Teor. Phys. {\bf 70}, 3 (1999)
[JETP Lett. {\bf 70}, 1 (1999)], hep-th/9905008; G. E. Volovik and
A. Vilenkin, Phys. Rev. D {\bf 62}, 025014 (2000), hep-ph/9905460.

\bibitem{14}
A. A. Andrianov, P. Giacconi, and R. Soldati, Grav. Cosmol. Suppl.
{\bf 8N1}, astro-ph/0111350; J. High Energy Phys.  02 (2002) 030.

\bibitem{9}
R. Jackiw and V. A. Kosteleck\'{y}, Phys. Rev. Lett. {\bf 82},
3572 (1999), hep-ph/9901358.

\bibitem{10}
M. P\'{e}rez-Victoria, Phys. Rev. Lett. {\bf 83}, 2518 (1999),
hep-th/9905061; JHEP {\bf 04}, 032 (2001).

\bibitem{11}
M. Chaichian, W. F. Chen and R. Gonzalez Felipe, Phys. Lett. B
{\bf 503}, 215 (2001), hep-th/0010129.

\bibitem{12}
J. M. Chung and P. Oh, Phys. Rev. D {\bf 60}, 067702 (1999),
hep-th/9812132; J. M. Chung, Phys. Rev. D {\bf 60}, 127901 (1999),
hep-th/9904037; J. M. Chung and B. K. Chung, Phys. Rev. D {\bf
63}, 105015 (2001), hep-th/0101097.

\bibitem{13}
W. F. Chen, Phys.Rev. D {\bf 60}, 085007 (1999), hep-th/9903258.

\bibitem{15} D. Ebert, V. Ch. Zhukovsky, and A. S. Razumovsky,
  Phys. Rev. D,  {\bf 70}, 025003, (2004),
  hep-th/0401241.

\bibitem {16} P. B. Pal and T. N. Pham,
Phys. Rev. D {\bf 40}, 259 (1989); J. F. Nieves,  Phys. Rev. D
{\bf 40}, 866 (1989); D. N\"{o}tzold and G. Raffelt, Nucl. Phys. B
{\bf 307}, 924 (1988); J. Pantaleone, Phys. Lett. B {\bf 268}, 227
(1991).

\bibitem{F51}
W. H. Furry, Phys. Rev. {\bf{81}}, 115 (1951).

\bibitem{klepikov}
N. P. Klepikov, Zh. Eksp. Teor. Fiz., {\bf 26}, 19 (1954). The
results of this pioneer work see, e.g., in A.~A.~Sokolov and
I.~M.~Ternov, {\emph{Radiation from Relativistic electrons}}
(American Institute of Physics Translations Series, New York,
1986).

\bibitem{adler}
S. L. Adler, Ann. Phys. (N.Y.), {\bf 67}, 599 (1971).

\bibitem{ritus}
A. I. Nikishov and V. I. Ritus, Zh. Eksp. Teor. Fiz. {\bf 46}, 776
(1964)[Sov. Phys. JETP {\bf 19}, 529 (1964)]; V. I. Ritus, Tr.
Fiz. Inst. Akad. Nauk SSSR {\bf 111}, 5 (1979) [J. Sov. Laser
Research {\bf 6}, 497 (1985)]; A. I. Nikishov, Tr. Fiz. Inst.
Akad. Nauk SSSR {\bf 111}, 152 (1979) [J. Sov. Laser Research {\bf
6}, 619 (1985)].

\bibitem{shabad}
I. A. Batalin and A. E. Shabad, Zh. Eksp. Teor. Fiz. {\bf 60}, 894
(1971) [Sov. Phys. JETP {\bf 33}, 483 (1971)];  A. E. Shabad, Ann.
Phys. (N.Y.) {\bf 90}, 166 (1975); Tr. Fiz. Inst. Akad. Nauk SSSR
{\bf 192}, 5 (1988) [A. E. Shabad, {\emph{Polarization of the
Vacuum and a Quantum Relativistic gas in an External Field}} (Nova
Science Publ., New York, 1991)]; A. E. Shabad and V. V. Usov,
Astrophys. Space Sci. {\bf 102}, 327 (1984).

\bibitem{L16}
I. M. Ternov, V. N. Rodionov, A. E. Lobanov, and O. F. Dorofeev,
Pis'ma Zh. Eksp. Teor. Fiz. {\bf 37}, 288 (1983) [Sov. Phys. JETP
Lett. {\bf 37}, 342 (1983)]; O. F. Dorofeev, V. N. Rodionov, and
I. M. Ternov, Pis'ma Zh. Eksp. Teor. Fiz. {\bf 40}, 159 (1984)
[Sov. Phys. JETP Lett. {\bf 40}, 917 (1984)]; I. M. Ternov, V. N.
Rodionov, O. F. Dorofeev, and A. E. Lobanov, {Izv. Vissh. Uchebn.
Zav., Fizika} {\bf 29}, No. 3, 82 (1986) [Sov. Phys. J. {\bf 29},
224 (1986)].

\bibitem{ZZ} I. M. Ternov, V. R. Khalilov, and V. N. Rodionov,
{\emph{Interaction of Charged Particles with a Strong
Electromagnetic Field}} (Moscow State University Press, Moscow,
1982) [in Russian].

\bibitem{kniga}
V. Ch. Zhukovskii, O. F. Dorofeev, and A. V. Borisov, in
{\emph{Synchrotron Radiation Theory and its Development, }}edited
by V. A. Bordovitsyn, Series on Synchrotron Radiation Technique
and Applications -- Vol. 5
 (World Scientific, Singapore, 1998),
pp. 350-400.

\bibitem{BVZ97}
A. V. Borisov, A. S. Vshivtsev, V. Ch. Zhukovsky, and P. A.
Eminov, Usp. Fiz. Nauk {\bf 167}, 241 (1997) [Phys. Usp. {\bf 40},
229 (1997)].

\bibitem{L05} A. E. Lobanov, Phys. Lett. B {\bf 619}, 136 (2005),
hep-ph/0506007.

\bibitem{ST}  A. A.~Sokolov and I. M. Ternov, {\emph{Synchrotron
radiation}}, (Akademie-Verlag, Berlin, 1968; Pergamon Press, N.Y.,
1968).

\bibitem{BG} V. G. Bagrov and D. M. Gitman, {\emph{Exact Solutions of
Relativistic Wave Equations}}, (Kluwer, Dordrecht, 1990).

\bibitem{T}  I. M. Ternov,  {\emph{Introduction to Physics of Spin of
Relativistic Particles}} [in Russian], (Moscow State Univ. Press,
Moscow, 1997).

\bibitem{x} Here, for convenience, we use the mathematical
notations: $\varnothing$ for the empty set, [a,b) for the open
interval, and [a,b] for the closed interval.

\bibitem{BE} H. Bateman and A. Erdelyi, {\emph{Higher transcendental
functions}}, v. 3, (McGraw-Hill Book Company, Inc., New
York--Toronto--London, 1955).

\bibitem{1e}
T. Jacobson, S. Liberati, D. Mattingly, and F. W. Stecker, Phys.
Rev. Lett. {\bf 93}, 021101 (2004), astro-ph/0309681.

\bibitem{BTB95} V. A. Bordovitsyn, I. M. Ternov, and V. G. Bagrov,
Usp. Fiz. Nauk {\bf 165}, 1084 (1995) [Phys. Usp. {\bf 38}, 1037
(1995)].

\bibitem{theor} See, e.g., I. M. Ternov, V. G. Bagrov, and
  R. A. Rzaev, Zh. Eksp. Teor. Fiz. {\bf 46}, 374 (1964) [Sov. Phys. JETP
{\bf 19}, 255 (1964)]; and for
  radiation of an electron with vacuum magnetic moment
  I. M. Ternov, V. G. Bagrov, and V. Ch. Zhukovsky, Vestnik
  Mosk. Univ., Fiz. Astron. {\bf 7}, No. 1, 30 (1966)
  [Moscow University Physics Bulletin {\bf 21}, No. 1, 21 (1966)].

\bibitem{exper}
S. A. Belomestnykh, A. E. Bondar, M. N. Yegorychev, V. N. Zhilich,
  G. A. Kornyukhin, S. A. Nikitin, et al., Nucl. Instr. \& Methods in
  Physics Research A, {\bf 227}, 173 (1984).

\bibitem {L49} A. Lobanov and A. Studenikin,
Phys. Lett. B {\bf 564}, 27 (2003), hep-ph/0212393.

\end{thebibliography}
\end{document}